\documentclass[conference]{IEEEtran}
\IEEEoverridecommandlockouts
\usepackage{cite}
\usepackage{amsmath,amssymb,amsfonts}
\usepackage{graphicx}
\usepackage{textcomp}
\usepackage{xcolor}
\usepackage[T1]{fontenc}

\usepackage{algorithm}
\usepackage{algpseudocode}
\usepackage{tabularx,ragged2e,siunitx}
\usepackage{diagbox}
\usepackage{multirow}
\usepackage{dblfloatfix}
\usepackage{booktabs}
\usepackage{subfig}

\usepackage[flushleft]{threeparttable}
\usepackage[normalem]{ulem}
\def\BibTeX{{\rm B\kern-.05em{\sc i\kern-.025em b}\kern-.08em
    T\kern-.1667em\lower.7ex\hbox{E}\kern-.125emX}}
\begin{document}

\newcommand\change[2]{#2}
\newcommand\changer{}

\title{Identifying Coordinated Activities on Online Social Networks Using Contrast Pattern Mining}

\author{\IEEEauthorblockN{Isura Manchanayaka}
\IEEEauthorblockA{\textit{School of Computing and Information Systems} \\
\textit{The University of Melbourne}\\
Melbourne, Australia \\
imanchanayak@student.unimelb}
\and
\IEEEauthorblockN{Zainab Zaidi}
\IEEEauthorblockA{\textit{School of Computing and Information Systems} \\
\textit{The University of Melbourne}\\
Melbourne, Australia \\
zainab.raziazaidi@unimelb.edu.au}
\and
\IEEEauthorblockN{Shanika Karunasekera}
\IEEEauthorblockA{\textit{School of Computing and Information Systems} \\
\textit{The University of Melbourne}\\
Melbourne, Australia \\
karus@unimelb.edu.au}
\and
\IEEEauthorblockN{Christopher Leckie}
\IEEEauthorblockA{\textit{School of Computing and Information Systems} \\
\textit{The University of Melbourne}\\
Melbourne, Australia \\
caleckie@unimelb.edu.au}
}

\maketitle

\begin{abstract}
    The proliferation of misinformation and disinformation on social media networks has become increasingly concerning. With a significant portion of the population using social media on a regular basis, there are growing efforts by malicious organizations to manipulate public opinion through coordinated campaigns. Current methods for identifying coordinated user accounts typically rely on either similarities in user behaviour, latent coordination in activity traces, or classification techniques. In our study, we propose a framework based on the hypothesis that coordinated users will demonstrate abnormal growth in their behavioural patterns over time relative to the wider population. Specifically, we utilize the EPClose algorithm to extract contrasting patterns of user behaviour during a time window of malicious activity, which we then compare to a historical time window. We evaluated the effectiveness of our approach using real-world data, and our results show a minimum increase of 10\% in the F1 score compared to existing approaches.
\end{abstract}


\section{Introduction}
\label{sec:introduction}
The surge in social media usage over the past decade can be attributed to various factors, including peer influence, the emergence of online communities, and the allure of following influential figures. However, this surge has also amplified the risks of misinformation and disinformation due to the ease of creating fake accounts. Particularly concerning is the role of coordinated behaviors in spreading misinformation and disinformation, often driven by political motives. Such campaigns require large numbers of user accounts working in coordination to propagate ideological agendas and achieve specific objectives.

The USA presidential election in 2016 was influenced by information operations carried out by Russia's Internet Research Agency (IRA) on Twitter and Facebook \cite{roberts.muellerUNITEDSTATESAMERICA2018}. The  Permanent Select Committee on Intelligence \cite{permanentselectcommitteeonintelligenceExposingRussiaEffort2018} identified 3,841 coordinated Twitter accounts and 470 Facebook pages that were affiliated with the IRA in 2017. In 2018, Twitter publicly released tweets and users related to this case. Additionally, in 2019, the UK general elections were influenced by coordinated users that polarized political opinions on Twitter \cite{nizzoliCoordinatedBehaviorSocial2021}.

Even though social media platforms claim to take measures to mitigate malicious campaigns by inspecting the behaviours of individuals, it is harder to identify a campaign as a whole due to its apparent natural growth and  organic behaviour. Hence, identifying coordination from a broader perspective is an important step to identify malicious campaigns. We believe that summarizing behavioural patterns within a community is a promising approach to achieve this broad perspective.


A promising approach for summarizing changes within datasets is pattern mining, a machine learnning technique well-established in domains such as medicine, education, and computer security. However, its application in social networks has thus far been limited, primarily focusing on bot detection. In this study, we leverage pattern mining to identify anomalous coordinating activities in online social networks.

The notion of patterns provides a means of encoding the behavioural patterns of users. Our hypothesis is that coordinated users will exhibit unusual correlated activity patterns that are not reflected among the wider population of users over time. For example, a user $u_1$ sharing a post from BBC News on a Monday morning can be represented by the following list of key, value pairs \{\emph{user\textnormal{: $u_1$, }is retweet?\textnormal{: yes, }original tweet's author\textnormal{: BBC News, }day of week\textnormal{: Monday, }time of day\textnormal{: 8 \textsc{am} - 10 \textsc{am}}}\}. If we observe this pattern over some time and there is an increase in the \change{support}{frequency} of that particular pattern, we can state that there is a growth in the \change{support}{frequency} of that pattern. If we can compare \change{support counts}{frequencies} of the behavioural patterns in the present-time behaviour with activity in an earlier reference time period that is assumed to be normal, we can interpret the growth of \change{support counts}{frequencies} compared to the background as anomalous behaviour. Such patterns can be identified as anomalous patterns. This method of comparing patterns from two sets of datasets is known as contrast pattern mining.


\change{In this paper, we focus on (1) investigating methods of identifying coordination using contrast pattern mining, (2) evaluating the performance of our model on real data, and (3) describing the nature of coordinated groups using the extracted patterns.}{}

Our experiments show that the social media users that are associated with contrasting behavioural patterns compared to their historical behaviours are likely to be coordinated in nature. We achieve F1 scores up to 86\% in identifying coordinating users for the IRA dataset, thus supporting our hypothesis. Moreover, the accuracy of our approach in terms of F1 score exceeds the corresponding accuracy of a range of benchmark approaches by more than 10\%.

We note the following as our contributions: (1) Formulating the usage of contrast pattern mining for identifying coordinated users, (2) Proposing a framework for making use of contrasting behavioural patterns for real data, (3) Conducting experiments comparing different parameters, attributes and approaches on real-life social network data to establish our claims.

In subsequent sections, we provide background and definitions (Section \ref{sec:preliminaries}), formulate our research problem (Section \ref{sec:problem-statement}), detail our methodology (Section \ref{sec:methodology}), present experiments, results, and analysis (Section \ref{sec:experiments}), and conclude with insights and future research directions (Section \ref{sec:conclusion}).

\section{Related Work}
\label{sec:related-work}
The ease of implementing and deploying inauthentic behaviors on social media has spurred research into various types of anomalous behaviors in online social networks (OSNs). Notably, coordinated efforts to spread misinformation and disinformation, often facilitated by social bots, have become prevalent \cite{cresciDecadeSocialBot2020}. Coordinating botnets emerged notably during the 2012-2013 period \cite{cresciDecadeSocialBot2020}. 

Several studies have endeavored to identify coordination in online social networks. Some focused on identifying campaigns in social media \cite{leeContentdrivenDetectionCampaigns2011, leeDetectingCollectiveAttention2012}, assuming coordination is reflected in message themes while overlooking other behavioral aspects. Network-based approaches \cite{pachecoUncoveringCoordinatedNetworks2021a, nizzoliCoordinatedBehaviorSocial2021, weberAmplifyingInfluenceCoordinated2021, magelinskiSynchronizedActionFramework2021, hristakievaSpreadPropagandaCoordinated2022} defined coordination in terms of community detection on user similarity graphs. \cite{weberAmplifyingInfluenceCoordinated2021} highlights strategies such as pollution, boost, and bully. However, network-based approaches perform well only when the networks are sparse enough \cite{magelinskiSynchronizedActionFramework2021}. In contrast, others define coordination based on the synchronicity of users over time \cite{zhangVigDetKnowledgeInformed2021, sharmaIdentifyingCoordinatedAccounts2021}, employing techniques like masked self-attention \cite{vaswani2017attention}.

Contrast pattern mining, also called emerging or discriminative pattern mining is a subfield within data mining that aims to uncover patterns that exhibit a significant difference in their frequency/support between two datasets. Contrast patterns reveal differences between two datasets while simultaneously summarizing the underlying patterns in datasets. Contrast pattern mining has been used in multiple applications, such as (1) medicine (discovering toxicological knowledge \cite{sherhodToxicologicalKnowledgeDiscovery2013, sherhodEmergingPatternMining2014}, predicting heart disease \cite{DBLP:books/crc/dong13/RyuLP13}), (2) education (student learning patterns \cite{kongAnalysisStudentsLearning2020, Thanasuan2017EmergingPI}), and (3) computer security (network anomaly detection \cite{alipourchavaryImprovingScalabilityContrast2020}, malware detection \cite{hellalMinimalContrastFrequent2016}). The use of contrast pattern mining in social media analysis has been limited to bot detection \cite{LoyolaGonzlez2019ContrastPC}, using a decision tree based approach to extract contrast patterns. We propose that contrast pattern mining is a promising approach to detecting coordination by identifying anomalous growth in activity patterns of subgroups of users. There have been a variety of approaches proposed for contrast pattern mining. \cite{dongEfficientMiningEmerging1999} first addressed the problem using a border-based method. Multiple decision tree-based approaches have also been proposed in the literature \cite{loyola-gonzalezPBC4cipNewContrast2017, canete-sifuentesClassificationBasedMultivariate2019}. One downside of decision tree-based approaches is that they are prone to overfitting. Methods based on frequent pattern trees (FP-Trees) such as EPClose \cite{alipourchavaryImprovingScalabilityContrast2020}, IEP-TFP \cite{piaoEnumerationTreeBased2011}, DPMiner \cite{Li2007MiningSI} have become popular due to their performance. In this paper, we build on the theory of EPClose to develop a method for detecting coordinated social network accounts.



\section{Preliminaries}
\label{sec:preliminaries}
\change{We adopt several definitions from \mbox{\citet{alipourchavaryImprovingScalabilityContrast2020}} and \mbox{\citet{dongContrastDataMining2016}}}{}
We refer to an interaction made by a user with the social network as an \emph{event}. An event can be stored using a list of (attribute, value) pairs. Values can be either numerical or categorical. The domain of values for an attribute $a$ is denoted $\Delta_a$. An \emph{item} is an (attribute, value) pair. A \emph{transaction} is a set of items. A transaction is associated with a \emph{transaction id}. For example, a single tweet is a transaction. In that case, the \texttt{tweetid} is the transacation id. The item (\texttt{username}, @abc) reflects the author of a tweet in a transaction. A list of such transactions is a \emph{transactional dataset}. \emph{Attribute space} $\mathcal{A}$ is the set of all attributes in a given dataset. A \emph{pattern} or an \emph{itemset} is a set of items. We say an itemset $X$ is contained in transaction $T$ iff. $X \subseteq T$. $f_D (X)$, the set of transactions that contain the pattern $X$ is defined as $\{T \in D \mid X \subseteq T\}$. The number of transactions in a dataset $D$ that contain pattern $X$ is the \emph{support count} of that pattern i.e., $SC(X,D)=|f_D (X)|$. The \emph{support} of a pattern $X$ is defined as $ supp(X, D) = \frac{SC(X, D)}{|D|}$. 

Patterns in the form of itemsets provide an opportunity to encode the behavioural patterns of users. An example of a behavioural pattern encoded as an itemset is, \{(\emph{user\textnormal{, $u_1$), (}is retweet?\textnormal{, yes), (}original tweet's author\textnormal{, BBC News), (}day of week\textnormal{, Monday), (}time of day\textnormal{, 8 \textsc{am} - 10 \textsc{am})}}\}. If we compare the support of such itemsets in a historical time window with the support of those itemsets in a subsequent time span containing anomalous activities, we can interpret the growth in support as anomalous behavioural patterns. Such itemsets with high growth can be identified as contrast patterns. The following are some key definitions related to contrast pattern mining.
The main dataset that is to be analysed and compared to other datasets is called the \emph{target dataset} $D_t$. A baseline dataset against which changes in the target dataset are found is called the \emph{background dataset} $D_b$. The \emph{growth rate} of a pattern is the ratio of its supports between the target and background datasets $gr(X,D_t,D_b )=\frac{supp(X,D_t)}{supp(X,D_b)}$. If $supp(X,D_b )=supp(X,D_t )=0$, then $gr(X,D_t,D_b )=0$ and if $supp(X,D_b )=0$ and $supp(X,D_t )>0$, then $gr(X,D_t,D_b )=\infty $. \emph{Support delta} is another way of measuring the growth of support of a given pattern, and is defined as $supp_\delta (X,D_t,D_b )=supp(X,D_t )-supp(X,D_b)$. A \emph{contrast pattern} $X$ is a pattern whose support in the target is significantly different from the background. Given a \emph{growth rate threshold} $\rho>1$ and a \emph{minimum support delta} $\sigma_\delta>0$, we say pattern $X$ is a contrast pattern iff $gr(X,D_t,D_b )\geq \rho$ or $supp_\delta (X,D_t,D_b ) \geq \sigma_\delta$. A contrast pattern $p$ takes the following form: $p=\{(a,v) \mid a\in \mathcal{A}, v \in \Delta_a\}$.

A pattern $X$ is called a \emph{closed pattern} iff there exists no superset $Y$ of $X$ satisfying $ SC(Y,D)=SC(X,D) $. A pattern $X$ is a \emph{closed contrast pattern} (CCP) iff $ supp(X,D_t )\geq\sigma>0, gr(X,D_t,D_b )\geq\rho>1 $ and $X$ is a closed pattern in $D_t \cup D_b $. Here, $\sigma$ is called the \emph{minimum support}.

\subsection{Example}

\begin{table}[H]
    \caption{Two example transaction tables. The column names are abbreviated as u -- user id, r -- is retweet?, ota -- original tweet's author}%
    \label{tab:example}%
    \centering
    \scriptsize
\subfloat[Background $D_b$]{
    \begin{tabular}{ccc}
        \toprule
        u & r & ota \\
        \midrule
        $u_2$ & no & - \\
        $u_1$ & yes & $u_2$ \\
        $u_1$ & no & - \\
        $u_2$ & yes & $u_3$ \\
        $u_2$ & no & - \\
        \bottomrule
    \end{tabular}
}
\quad
\subfloat[Target $D_t$]{
    \begin{tabular}{ccc}
        \toprule
        u & r & ota \\
        \midrule
        $u_1$ & yes & $u_2$ \\
        $u_1$ & yes & $u_2$ \\
        $u_2$ & no & - \\
        $u_1$ & yes & $u_2$ \\
        $u_4$ & no & - \\
        \bottomrule
    \end{tabular}
}
\end{table}
This section gives an example for each definition in the text above. An example dataset is split into background and target partitions as shown in Table \ref{tab:example}. The set $T_1=\{$(u, $u_2$), (r, no), (ota, -)$\}$ is the first transaction of $D_b$, which consists of three attribute-value pairs. Any subset of a transaction is a pattern/itemset. If we examine the pattern $p_0=\{$(u, $u_1$), (r, yes), (ota, $u_2$)$\}$, it is apparent that there is only one instance of that pattern in $D_b$. Hence, $SC(p_0, D_b)=1$. Similarly, $SC(p_0, D_t)=3$. It follows that $supp(p_0, D_b)=\tfrac{1}{5}$ and $supp(p_0, D_b)=\tfrac{3}{5}$. The growth rate and support delta can be calculated as $gr(p_0, D_t, D_b)=\frac{3/5}{1/5}=3, supp_{\delta}(p_0, D_t, D_b)=\tfrac{3}{5}-\tfrac{1}{5}=\tfrac{2}{5}$. Pattern $p_0$ is a closed pattern since there is no superset of $p_0$ with a higher support count in any of $D_b$ and $D_t$. Hence, in a scenario of a minimum support ($\sigma$) of 2 and a growth rate threshold ($\rho$) of 1.5, $p_0$ can be classified as a closed contrast pattern.

\section{Problem Statement}
\label{sec:problem-statement}
The challenge we address is how to identify anomalous coordinting behaviour among social media users in a robust manner. This requires a novel approach to identify a combination of features from user posts that succinctly characterise the change in behaviour in contrast to normal behaviour. To address this challenge, we build upon the theory of contrast pattern mining, which provides a robust and scalable method for exploring the huge search space of possible feature combinations.

Let $D$ be a set of posts in an online social network \change{}{in a time interval $[t_s, t_e]$}. \change{The set $D$ is compiled such that (1) every element is a post (e.g., a tweet on Twitter) authored by one of a pre-determined set of users $U$ and, (2) every post in $D$ is created on the internet in a pre-determined time interval $[t_s, t_e]$.}{} Each element of the set $D$ takes the form of a transaction. Say we determine two time intervals $[t_0, t_1]$ and $[t_2, t_3]$ such that $t_s \leq t_0 < t_1 \ll t_2 < t_3 \leq t_e$,  $[t_2, t_3]$ presumably contains anomalous activities based on observations, and $[t_0, t_1]$ presumably does not contain anomalous activities. Let $D_b$ and $D_t$ be the subsets of $D$ such that each post in $D_b$ is created in the time interval $[t_0, t_1]$ and each post in $D_t$ is created in the time interval $[t_2, t_3]$. \change{It is preferable for $t_1$ and $t_2$ to be as far apart as possible since it is likely to contain similar behavioural patterns in closer time spans. }{} \change{With the preliminaries above and the partitioned data, we formulate the studied problem statement as follows.}{}

\noindent\textbf{%
Problem Definition.
}\emph{Given two transactional datasets of posts ($D_b$ and $D_t$), find the behavioural patterns that have a significant growth from a \change{defined}{} historical time span $[t_0, t_1]$ that is likely to have few anomalous coordinating behaviours to a \change{defined}{} subsequent time span $[t_2, t_3]$ that is highly likely to have anomalous coordinating behaviours. Find the users that are associated with such behavioural patterns and test the hypothesis that: users who show anomalous behaviour are likely to be associated with contrasting behavioural patterns compared to their historical behaviour and other normal users.}

\section{Methodology}
\label{sec:methodology}

\begin{figure*}[t!]
    \centering
    \includegraphics[width=0.8\textwidth]{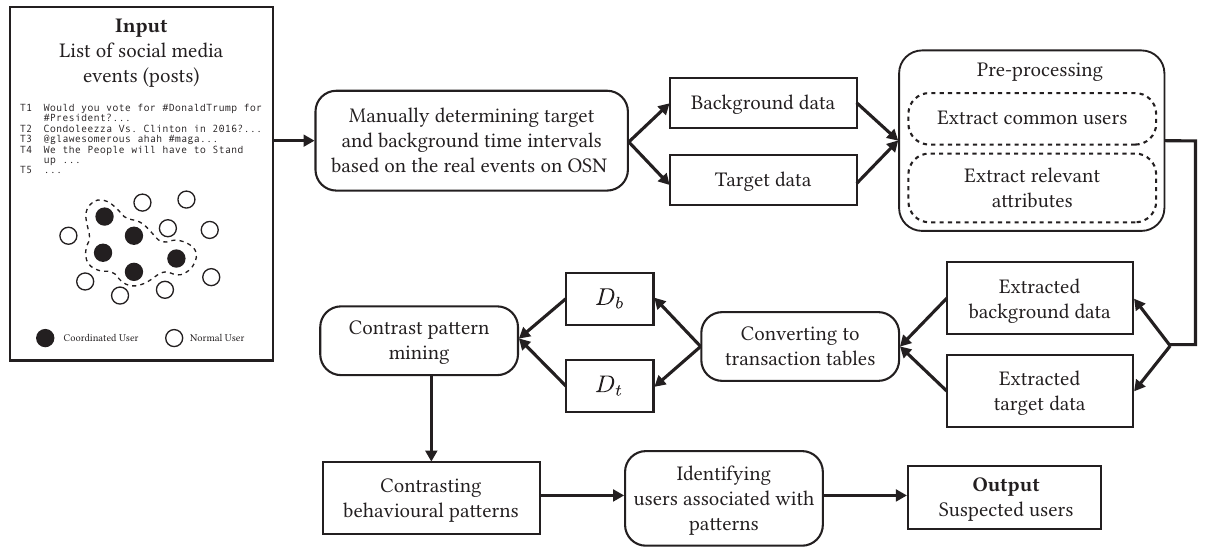}
    \caption{Overview of the proposed framework to identify suspected coordinated user accounts based on contrast pattern mining. Boxes with rounded corners represent a process while rectangular boxes represent data. The dark circles denote coordinating users and empty circles denote normal users.}
    \label{fig:flow}
\end{figure*}

This section outlines the proposed framework to solve the problem we identified above. Initially, we determine the background and target time intervals based on observations of the dataset of posts. Subsequently, we pre-process the data to extract relevant attributes and users. We then apply a contrast pattern mining algorithm to the converted transaction tables. Finally, we extract suspicious users by using the contrast patterns that we have obtained. An overview of our framework is presented in Figure 1.

\subsection{Overview}
\label{overview}

Given background and target transactional datasets, we derive the set of contrast patterns $\mathcal{P}$ using a contrast pattern mining algorithm. Given a set of attributes $A\subseteq \mathcal{A}$, the subset of $\mathcal{P}$ where each contrast pattern is associated with every attribute in $A$ is the set of filtered contrast patterns $\mathcal{P}_A$. Define $attribs(p)=\{a \mid a \in \mathcal{A}, (a,v)\in p \}$, i.e., the set of attributes of a given contrast pattern $p$. Then, $\mathcal{P}_A=\{p \mid A \subseteq attribs(p)\}$. $\mathcal{P}_A$ can be obtained by filtering patterns in $\mathcal{P}$, which only contains all attributes in $A$. Since $\mathcal{P}_{\{user\}}$ contains the contrast patterns with emerging behavioural patterns for users, we claim that the users that appear in  $\mathcal{P}_{\{user\}}$ are the set of suspicious of users. The set of users $U_{suspicious}$ that appear in $\mathcal{P}_{\{user\}}$ is marked as identified anomalous coordinating users. Formally, $U_{suspicious} = \{u \mid p \in \mathcal{P}_{\{user\}},  (user, u) \in p \}$. \change{For a Twitter dataset, the attribute $user$ and \texttt{userid} are identical.} Even though there are multiple parameters in the algorithm, our experiments suggest that the performance of our model is mainly affected by the $\rho$ and $\sigma$ parameters.

\subsection{Partitioning}
Given a dataset, the time interval for $D_t$ should be the period of time when we suspect the existence of anomalous behaviours. The time interval $D_b$ should be a historical time period when the activity of users in $D_t$ is sufficient. It is necessary for these two time periods to be as far apart as possible such that there is no obvious influence between them.

\subsection{Pre-processing}
\label{preprocessing}

\begin{itemize}
    \item\label{subsubsec:extracting-common-users} \noindent\emph{%
    Extracting common users.
    }  For a user to contribute to a contrast pattern, that user must be active in both $D_b$ and $D_t$. Hence, common users are extracted for the sake of efficiency.

    \item \noindent\emph{%
    Extracting relevant attributes.
    }  Categorical values are required for the purpose of grouping similar itemsets due to the low likelihood of matching real-valued observations for equality in practice. Binning or other pre-processing steps are needed for numerical values. For a Twitter dataset, the following fields were selected such that the attribute space consists of categorical values only -- user id, user reported location, tweet language, tweet time – divided into two fields; day of week and time of day (12 equal sized time slots per day), tweet client, is the tweet a retweet?, author of the original tweet if retweeted, list of segmented hashtags - each hashtag segmented using a Twitter corpora, and list of user mentions. The fields in the form of a list, such as hashtags and user mentions, were flattened out in transactions.

    \item \noindent\emph{%
    Converting to transactional datasets.
    } 
    \change{At this stage, the data is in vector format. }Posts in the extracted datasets are converted to lists of (attribute, value) pairs, i.e., transactions. For multivalued attributes\change{(e.g., hashtags, user mentions in Twitter)}{},  an itemset is generated with the same attribute. \change{Each (attribute, value) pair is then encoded with a unique integer identifier}{} \change{for the contrast pattern mining algorithm. At the end, we are left with two lists of transactions for the background and target datasets.}{}
\end{itemize}

\subsection{Mining coordinated users}
\label{mining}

\begin{itemize}
    \item \noindent\emph{%
    Contrast pattern mining.
    }  \change{FPClose \cite{Grahne2005FastAF} is a depth-first algorithm for contrast pattern mining that uses FP-trees for the purpose of extracting closed patterns. }{}EPClose \cite{alipourchavaryImprovingScalabilityContrast2020} is a \change{}{fast} scalable algorithm that extracts closed contrast patterns during closed pattern generation. \change{We leverage the promising performance of EPClose in order to extract contrast patterns.
    Instead of applying a threshold to $supp(X, D_t)$, we apply the threshold to $SC(X, D_b)$. Here onwards, this article refers to that threshold as the \emph{minimum support} with the symbol $\sigma$.}{} It should be noted that applying a threshold to support count ($SC$) is equivalent to applying a corresponding threshold to support ($supp$) since $supp \propto SC$. By applying the threshold to $D_t$, the algorithm outputs patterns whose support count in $D_b$ is $0$, which is unexpected since we are interested in patterns whose support is growing from a non-zero value in $D_b$ to $D_t$. Thus, we modify the threshold to support in $D_b$ instead of $D_t$. \change{}{Here onwards, this article refers to that threshold as the \emph{minimum support} with the symbol $\sigma$.}

    \item \noindent\emph{%
    Identifying users associated with patterns.
    }  The users that appear in contrast patterns are extracted and marked as coordinating users. $\mathcal{P}_{\{user\}}$ is constructed using $\mathcal{P}$ from the last step. \change{Then the set of users that appear in $\mathcal{P}_{\{user\}}$ is extracted. Occasionally, the contrast pattern mining algorithm may generate patterns of length one. Such patterns are omitted since they do not reveal an informative behavioural pattern.}{}
\end{itemize}

\section{Experiments}
\label{sec:experiments}

\subsection{Data}
\begin{figure}
  \centering
  \includegraphics[width=0.8\linewidth]{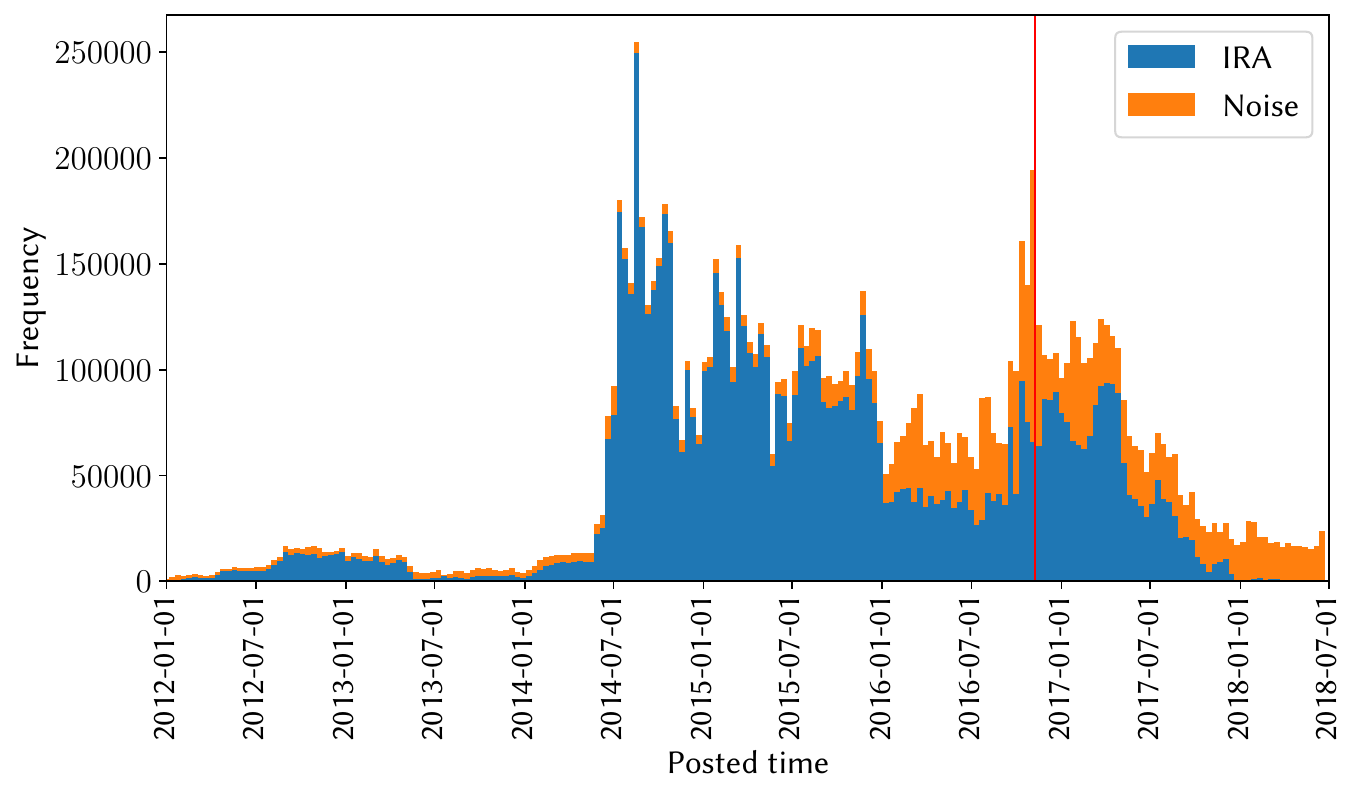}
  \caption{Stacked distribution of IRA activities and extracted noise tweets    across time. The bin size for the x-axis is 1 million seconds ({\raise.17ex\hbox{$\scriptstyle\mathtt{\sim}$}}11.6 days). The red vertical line shows the election date.}
  \label{fig:data-distribution}
\end{figure}

We experiment on the dataset of the activity of Russia's Internet Research Agency (IRA) influencing the 2016 USA presidential elections \cite{permanentselectcommitteeonintelligenceExposingRussiaEffort2018, roberts.muellerUNITEDSTATESAMERICA2018}, which consists of confirmed coordinated activities. This is a widely used dataset for detecting coordination \cite{weberAmplifyingInfluenceCoordinated2021,sharmaIdentifyingCoordinatedAccounts2021, zhangVigDetKnowledgeInformed2021,weberTemporalNuancesCoordination2022} since it is the only dataset with ground truth information. The dataset consists of 8.76 million tweets posted by 3613 users. \change{The dataset originally consisted of the following fields; Tweet id, User id, User display name, User screen name, User reported location, User profile description, User profile url, Follower count, Following count, Account creation date, Account language, Tweet language, Tweet text, Tweet time, Tweet client name, Replied tweet id, Replied user id, Quoted tweet id, Whether the tweet is a retweet, Retweeted user id, Retweeted tweet id, Latitude where the tweet is posted, Longitude where the tweet is posted, Quote count, Reply count, Like count, Retweet count, List of hashtags, List of urls, List of user mentions, List of poll choices if the tweet includes a poll.}{} Figure \ref{fig:data-distribution} shows the distribution of activity across the time.


In order to test the effectiveness of a coordination detection model, we introduce a set of noisy background events to the IRA dataset, since the IRA dataset only contains the set of coordinating users. \change{For that purpose, we scraped Twitter data for that period of time which includes the same popular hashtags in the IRA dataset using the Twitter API v2 for academics.}{} The criteria that were used to extract noise data were: posted time between 2008 and 2018, marked location anywhere in the USA, contains either one of the following hashtags - \emph{Election2016}, \emph{MAGA}, \emph{MakeAmericaGreatAgain}, \emph{AmericaFirst, DonaldTrump}, \emph{WakeUpUSA}, \emph{Trump}, \emph{TrumpTrain}, \emph{HilaryClinton, Trump2016}, \emph{DrainTheSwamp}, \emph{TrumpPence16}, \emph{tcot}, \emph{POTUS}, \emph{GOP}, \emph{Resist}, \emph{UniteBlue}, \emph{NeverHillary}, \emph{ElizabethWarren}, \emph{WeThePeople}, \emph{IllegalAliens}, \emph{TrumpRussia, ImWithHer}, \emph{GayHillary}, \emph{WakeUpAmerica}. The above set of hashtags were the top-occurring hashtags in the original IRA dataset. \change{This is to ensure that the noise data belongs to the same ongoing discussions at that period of time. }{}The background data of normal users consists of 2.80 million tweets from 333 thousand of users. \change{The distribution of coordinating tweets and the noisy tweets are shown in Figure \ref{fig:data-distribution}.} \change{High activity is apparent near the election time period (November 2016).}{}

\subsection{Experimental Setup}

\subsubsection{Time intervals}\label{subsubsec:baselines} Dataset $D_t$ was chosen such that it includes the election time period (November 2016). The range of the posting times of tweets in $D_t$ was the period of four months ending in November 2016. The time interval for the dataset $D_b$ was the period of four months ending in May 2015. This is entirely based on observations.

\subsubsection{Extracting top users} In addition to the step described in Section \ref{subsubsec:extracting-common-users}, for the purpose of evaluation, we further extract the top $n_C$ number of coordinating users and \change{}{top} $n_N$ number of normal users from the common users from the background and target datasets \change{}{using posting frequency} for the sake of performance. \change{Step \ref{subsubsec:extracting-common-users} is replaced with Algorithm \ref{alg:filter-top}, where $n_C$ and $n_N$ were chosen as 400 in the presented results.}{For the presented results, $n_C$ and $n_N$ were chosen as 400.}


\subsubsection{Baselines} We choose the following baselines in order to compare our results.
\begin{enumerate}
    \item \emph{Tweet frequency growth.} Instead of counting patterns and comparing supports, we formulate a comparison between the frequencies of tweets of each user in order to verify that our results are not due to the general growth of tweeting frequency that we see in the tweet density plots. Say the frequency of a user $u$ posting in dataset $D$ is $freq(u, D)$. Then for a given $\sigma$ and $\rho$\change{(building a similar analogy to contrast patterns)}{}, we can check the following requirements; (a) $freq(u, D_b) \geq \sigma > 0$. (b) Say $g(u, D_b, D_t)=\frac{freq(u, D_t) / |D_t|}{freq(u, D_b) / |D_b|}$ and $g(u, D_b, D_t) \geq \rho$ > 1. If (a) and (b) satisfies, then we mark them as suspecting users.
    \item \emph{Tweet language.} Since most (82\%) of the data in the coordinated set of users are in Russian and most (93\%) of the data in the noise data are in English, we compare our results with the results of a model that only use the language to determine the coordinated status. This model simply classifies a user to be coordinated if the language is Russian.
    \item \emph{LCN+HCC \cite{weberAmplifyingInfluenceCoordinated2021}.} This approach aims to identify coordinated communities using community detection on user similarity (retweet, mention, hashtags equally weighted) graphs. The temporal aspect is considered by a windowing mechanism. For a fair comparison, we use the dataset $D_b + D_t$ as the input. We use the window size as 10 days as the window length parameter and the threshold for FSA\_V to be 0.3.
    \item \emph{QT-LAMP-EP-BH \cite{komiyamaStatisticalEmergingPattern2017}.} Replace EPClose by the above contrast pattern algorithm. \change{They statistically guarantee the extracted contrast patterns to be more fair and accurate than other approaches. }{}In QT-LAMP-EP-BH, false discovery rates (FDR) are controlled using Benjamini-Hochberg (BH) method.
    \item \emph{QT-LAMP-EP-BY \cite{komiyamaStatisticalEmergingPattern2017}.} Instead of BH, use Benjamini-Yekutieli (BY) method to control FDRs.
    \item \emph{AMDN-HAGE.} \cite{sharmaIdentifyingCoordinatedAccounts2021} The SOTA for identifying coordinated users. We use the same set of hyperparameters except the threshold to determine the output influence values. Instead, we maximize the F1 score to determine it.
\end{enumerate}


\begin{figure*}
	\centering 
                \includegraphics[width=0.785\linewidth]{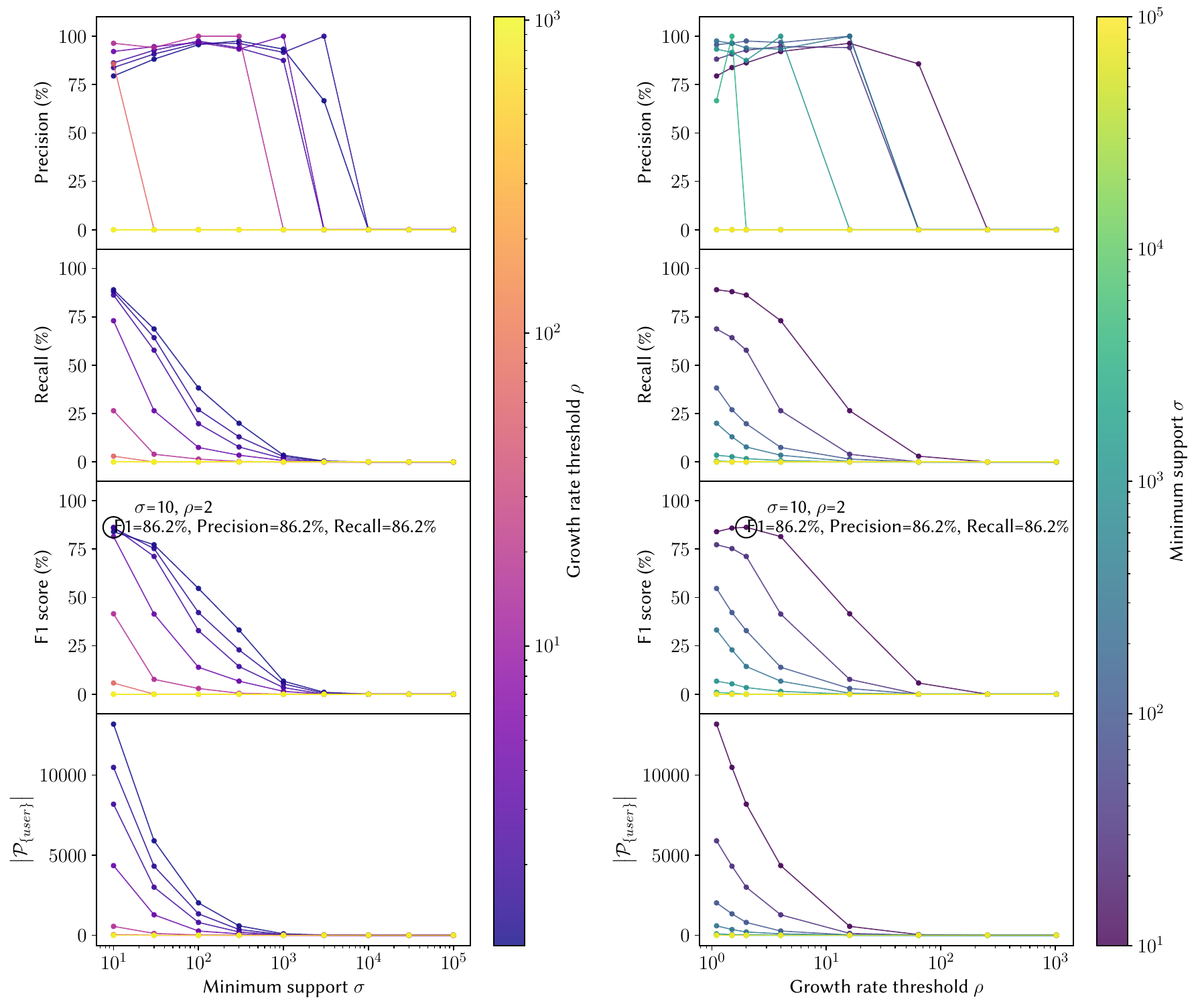}
        \caption{The variation of precision, recall, F1 score and the number of contrast patterns that are associated with users ($\left|\mathcal{P}_{\{user\}}\right|$) with the variation of $\sigma$ (left) and $\rho$ (right). Results with the maximum F1 score are circled in the 3rd row. Target time period: 2016/07 - 2016/11. Background time period: 2015/01 - 2015/05.}
        \label{fig:4}
\end{figure*}

\subsection{Results}

\subsubsection{
Quantitative Results
} The variation of precision, recall, F1 score and the number of contrast patterns associated with users with respect to variation in the growth rate threshold ($\rho$) and minimum support ($\sigma$) is shown in Figure \ref{fig:4}. The parameters and the performance metrics corresponding to the experiment with the highest F1 score is circled. There is a general increase in precision when $\sigma$ and $\rho$ are increased until a failure point. By increasing $\sigma$ and $\rho$, less relevant contrast patterns get filtered out. Promising precision values prove that the users that are associated with contrast patterns are likely to be coordinating users, thus supporting our statement in the problem definition. For large enough $\sigma$ and $\rho$, there are no contrast patterns, to say nothing of contrast patterns that are associated with users. Hence, the precision and recall is zero after some failure point. When we filter out more and more contrast patterns, the chance of us leaving out more and more relevant users is increased. Thus, the recall keeps decreasing when $\sigma$ and $\rho$ is increased. F1 score demonstrates the balance between the precision and recall, and it should be noted that F1 score is maximized for low $\sigma$ and $\rho$ values. We recommend that $\sigma=10$ and $\rho \in [1.1, 2]$ are parameters that yield generally good performance.

\subsubsection{
Baseline Comparisons
} A comparison between our approach and the baselines in Section \ref{subsubsec:baselines} is shown in Table \ref{tab:compare}.  Multiple experiments were carried out for each dataset by changing their parameters. The model with the maximum F1 score is displayed in the table. The performance of our model is satisfactory for both small and large datasets compared to the other approaches. The \emph{tweet frequency} baseline reveals that our results are not due to a general growth of frequency in tweets by each user. The missing cells are due to the heavy resource usage of the QT-LAMP-EP-* methods. The dataset $n_C=n_N=200$ takes {\raise.17ex\hbox{$\scriptstyle\mathtt{\sim}$}}80 GB of memory for those baseline methods. In contrast, EPClose needs {\raise.17ex\hbox{$\scriptstyle\mathtt{\sim}$}}300 MB of memory for contrast pattern mining for the case of $n_C=n_N=400$. The memory usage and the results for that case demonstrate the scalability of our model.

\begin{table*}
  \centering
    \caption{Results for detecting coordinated users using different methods. $n_C$ - number of coordinating users in the dataset, $n_N$ - number of normal users in the dataset.}
    \label{tab:compare}
    \begin{tabular}{lccccccccc}
      \toprule
      \multirow{2}{*}{Method} & \multicolumn{3}{c}{$n_C=n_N=100$} & \multicolumn{3}{c}{$n_C=n_N=200$} & \multicolumn{3}{c}{$n_C=n_N=400$} \\
      \cmidrule(lr){2-4} \cmidrule(lr){5-7} \cmidrule(l){8-10} 
      & Precision & Recall & F1 Score & Precision & Recall & F1 Score & Precision & Recall & F1 Score \\
      \midrule
      Tweet frequency & 92.1\% & 35.0\% & 50.7\% & 98.0\% & 48.0\% & 64.4\% & 91.5\% & 43.3\% & 58.7\% \\
      Tweet language & 64.0\% & 80.0\% & 71.1\% & 66.0\% & 81.0\% & 72.7\% & 66.0\% & 75.0\% & 70.2\% \\
      QT-LAMP-EP(BH) & 59.8\% & 58.0\% & 58.9\% & 84.8\% & 64.0\% & 72.9\% & - & - & - \\
      QT-LAMP-EP(BY) & 59.1\% & 55.0\% & 57.0\% & 85.1\% & 63.0\% & 72.4\% & - & - & - \\
      LCN+HCC & 76.1\% & 63.0\% & 68.9\% & 77.3\% & 65.0\% & 70.6\% & 81.5\% & 70.4\% & 75.5\% \\
      AMDN-HAGE & 50.0\% & 98.0\% & 66.2\% & 50.4\% & 100\% & 67.0\% & 50.6\% & 100\% & 67.2\% \\
      \hline
      Our approach & 77.1\% & 81.0\% & \textbf{79.0\%} & 88.6\% & 82.0\% & \textbf{85.2\%} & 86.2\% & 86.2\% & \textbf{86.2\%} \\
      \bottomrule
    \end{tabular}
  \end{table*}

  \subsubsection{
    Analysis of the Identified Patterns
  } 
\begin{figure}
    \centering
    \includegraphics[width=\linewidth]{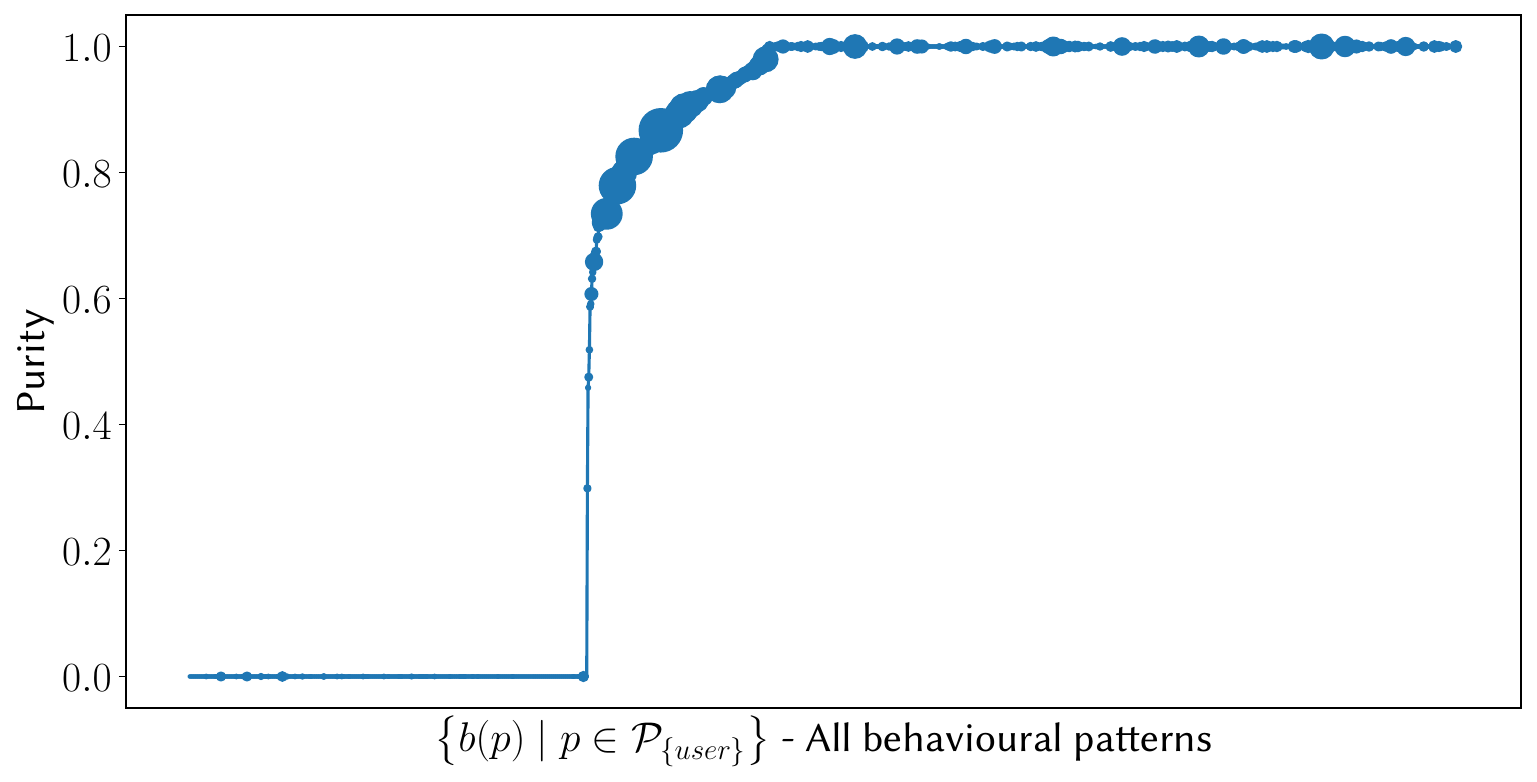}
    \caption{The distribution of purity values for each identified behavioural pattern. The size of each marker is proportional to the number of users associated with each behavioural pattern.}
    \label{fig:purity}
  \end{figure}

We selected the experiment that yielded the best F1 score from the previous section and performed an analysis on the identified patterns. Each identified pattern $p\in\mathcal{P}_{\{user\}}$ is associated with a $user$. The set of items in $p$ except the user-item corresponds to a behvaioural pattern $b(p)=\left\{(a,v) \in p \mid a \neq user \right\}$. Even though $p$ is uniquely associated with a user, it should be evident that the itemset $b(p)$ could be shared among multiple users. If $b(p)$ is shared only within the set of coordinated users, we say $p$ is a \textit{purely coordinated pattern}. If $b(p)$ is shared only within the set of normal users, we say $p$ is a \textit{purely normal pattern}. We call the rest of the patterns as \textit{mixed class patterns}. Given the set of known coordinated users $U_C$, then the target dataset of posts of coordinating users can be derived as $D_{t, C}=\{T \in D_t \mid (user, u) \in T, u \in U_C \}$. Using a similar approach to the concept of attack ratio in \cite{alipourchavaryImprovingScalabilityContrast2020}, we formulate a measure of purity for a contrast pattern $p$ as $purity(p)=\frac{SC(b(p), D_{t, C})}{SC(b(p), D_t)}$. When the purity is $1$, $p$ can be interpreted as matching purely coordinated users. When the purity is $0$, $p$ matches purely normal users. Figure \ref{fig:purity} shows the purity for each of the identified contrast patterns. Since $b(p) \subset p$, each $p$ can be grouped by $b(p)$ along the x-axis. For this dataset, it is apparent that there is a clear separation between the two classes. It shows the distinctive behavioural patterns of the coordinating and normal users. Table \ref{tab:pure-coordinated} and \ref{tab:pure-normal} shows a selected set of purely attacking and purely normal behavioural patterns respectively.
  
  \renewcommand*{\thefootnote}{\fnsymbol{footnote}}
  \newcolumntype{L}[1]{>{\raggedright\let\newline\\\arraybackslash\hspace{0pt}}m{#1}}
  \newcolumntype{C}[1]{>{\centering\let\newline\\\arraybackslash\hspace{0pt}}m{#1}}

  \newcommand\field[1]{\texttt{\textcolor{gray}{#1}}}
  
  \begin{table*}
    \caption{A selected set of purely coordinated behavioural patterns. User ids are censored.}
    \label{tab:pure-coordinated}
    \begin{threeparttable}
      \begin{tabular}{>{\hangindent=2em}L{13cm} C{1.5cm} C{1.5cm} }
      \toprule
      Behavioural pattern & Number of posts in $D_t$ & Number of users \\
      \midrule
      \field{tweet\_language}: en, \field{is\_retweet}: False, \field{hashtag}: news & 9536 & 3 \\
      \field{hashtag}: politics, \field{tweet\_language}: en, \field{is\_retweet}: False & 2181 & 5 \\
  \field{time\_of\_day}: 12 \textsc{pm}-14 \textsc{pm}\text, \field{is\_retweet}: False, \field{day\_of\_week}: \textsc{Monday}, \field{tweet\_language}: ru & 389 & 12 \\
  \field{user\_mentions}: 102\ldots, \field{tweet\_language}: ru & 252 & 2 \\
  \field{user\_mentions}: 124\ldots, \field{is\_retweet}: True, \field{retweet\_userid}: 124\ldots, \field{tweet\_language}: ru & 41 & 2 \\
  \field{hashtag}: Russian*, \field{hashtag}: Putin*, \field{is\_retweet}: False & 32 & 2 \\
  \field{time\_of\_day}: \textsc{18 pm - 20 pm}, \field{day\_of\_week}: \textsc{Wednesday}, \field{hashtag}: St. Petersburg*, \field{hashtag}: Nevsky News*,  \field{is\_retweet}: False & 31 & 1 \\
      \bottomrule
    \end{tabular}
  
    \begin{tablenotes}
      \item[*] Translated from Russian to English.
      \end{tablenotes}
  \end{threeparttable}
  \end{table*}

  \begin{table*}
    \caption{A selected set of purely normal behavioural patterns.}
    \label{tab:pure-normal}
    \begin{tabular}{>{\hangindent=2em}L{13cm} C{1.5cm} C{1.5cm} }
      \toprule
      Behavioural pattern & Number of posts in $D_t$ & Number of users \\
      \midrule
      \field{hashtag}: tcot, \field{tweet\_language}: en, \field{is\_retweet}: False & 22127 & 17 \\
      \field{hashtag}: tcot, \field{tweet\_language}: en, \field{is\_retweet}: False, \field{time\_of\_day}: \textsc{16 pm - 18 pm} & 1912 & 3 \\
      \field{hashtag}: rednationrising, \field{is\_retweet}: False & 687 & 2 \\
      \field{hashtag}: gop, \field{tweet\_language}: en, \field{is\_retweet}: False, \field{day\_of\_week}: \textsc{Thursday} & 124 & 2 \\
      \field{hashtag}: wakeupamerica, \field{tweet\_language}: en, \field{is\_retweet}: False & 120 & 3 \\
      \field{hashtag}: teaparty, \field{tweet\_language}: en, \field{is\_retweet}: False & 80 & 3 \\
      \field{day\_of\_week}: \textsc{Wednesday}, \field{tweet\_language}: en, \field{is\_retweet}: False, \field{hashtag}: tgdn & 68 & 2 \\
      \bottomrule
    \end{tabular}
  \end{table*}

\subsubsection{Ablation Study}
The impact of the choice of attributes was tested using a greedy approach. We searched the impact of attributes using two methods.
\begin{enumerate}
    \item \emph{Subtractive.} We start from attribute set $A_0=\mathcal{A}$. For each attribute $a$ that is in $A_i$ except \texttt{userid}, we remove that attribute, and we test multiple $\sigma$ and $\rho$ values to select the model with the maximum F1 score. Then we determine the attribute $a$ that results in the model with the minimum of that maximum F1 score to be the attribute that has the greatest impact at the stage of $A_i$. Hence, $A_{i+1}$ is created by removing that attribute from $A_i$. Define $F1(\sigma, \rho, A)$ to be the F1 score of the model with the attribute set $A$, minimum support $\sigma$, and growth rate threshold $\rho$. 
    We end the procedure when $A_n=\{\texttt{userid}\}$ where $n=|\mathcal{A}|-1$.
    \item \emph{Additive.} We start from attribute set $A_0=\{\texttt{userid}\}$. For each attribute $a$ that is not in $A_i$, we test multiple $\sigma$ and $\rho$ values to select the model with the maximum F1 score. Then we determine the attribute $a$ that results in that model with the maximum of that maximum F1 score to be the attribute that has the greatest impact at the stage of $A_i$. Hence, $A_{i+1}$ is created by appending that attribute to $A_i$.  
    We end the procedure when $A_n=\mathcal{A}$ where $n=|\mathcal{A}|-1$.
\end{enumerate}

\begin{figure*}
	\centering
        \includegraphics[width=0.9\linewidth]{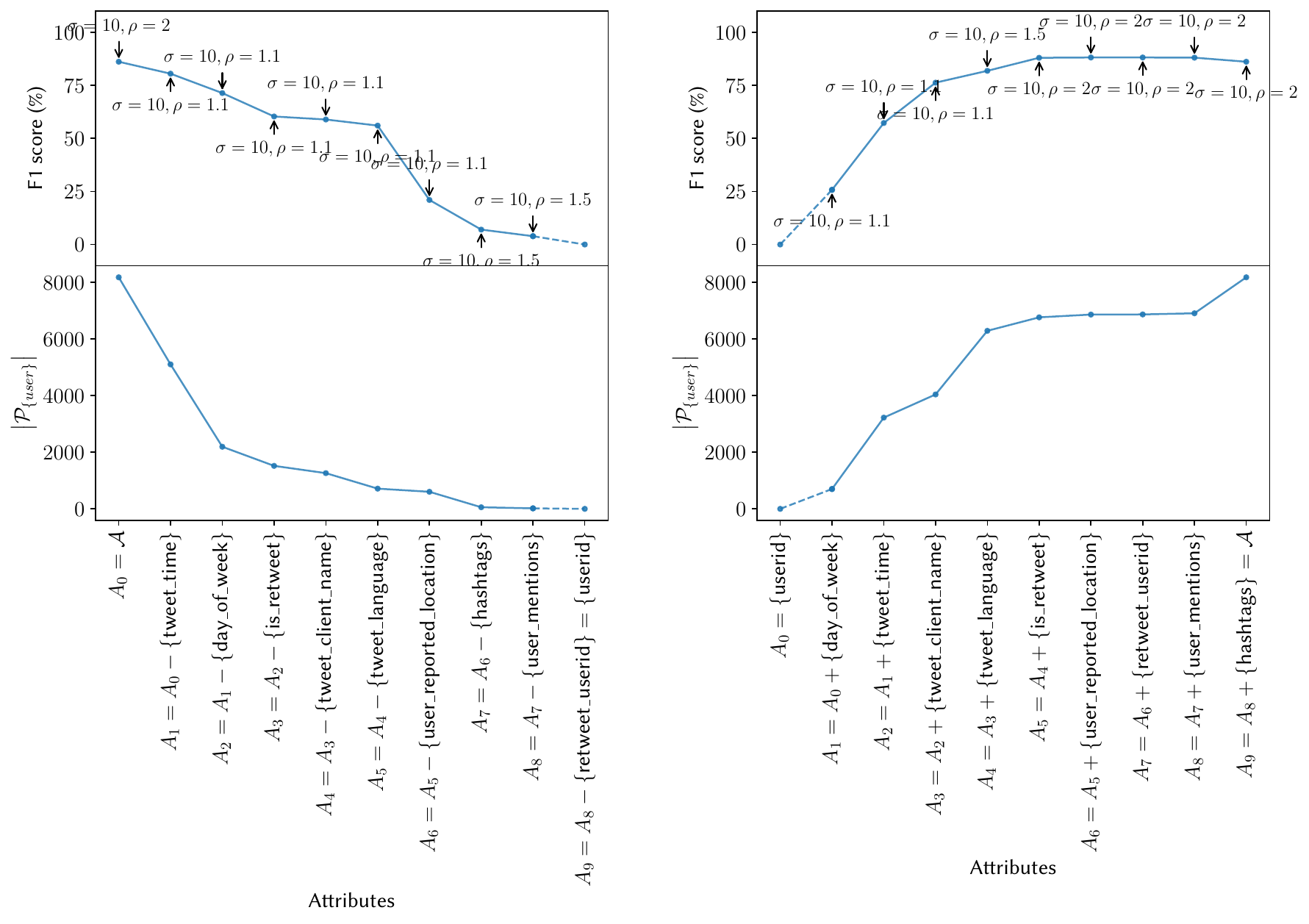}
        \caption{The variation of F1 scores and the number of contrast patterns that are associated with users ($\left|\mathcal{P}_{\{user\}}\right|$) when the highest impacting attribute is removed (left) or added (right) to the attribute set. $n_C=n_N=400$.}
        \label{fig:impact}
\end{figure*}

The purpose of the above methods is to search the locally optimal attribute sets and thus to assess the order of importance for each attribute. Figure \ref{fig:impact} shows the variation of F1 scores and the number of contrast patterns when the highest impacting attribute is removed or introduced to the existing attribute set. The subtractive method reveals that \texttt{tweet\_time}, \texttt{day\_of\_week}, \texttt{is\_retweet} are the three highest impact attributes in that order. The additive method reveals that \texttt{day\_of\_week}, \texttt{tweet\_time}, \texttt{tweet\_client\_name} are the three highest impact attributes in that order. Both methods identify that the first two attributes are most important compared to the rest of the attributes. Thus, temporal aspects of the events have played a major role in the behavioural patterns of this set of coordinating users. It is interesting to observe this indication of automation. Further, it is apparent that even for low numbers of attributes, similar to all attribute cases, low $\sigma$ and $\rho$ values yield the best possible F1 scores. 

\section{Conclusion}
\label{sec:conclusion}
In this work, we proposed a novel approach to identify coordination by exploiting the growth of behavioural patterns of users in an online social network. We demonstrate that our proposed approach can achieve an increase of at least 10\% in the F1 score compared to existing approaches. It should be noted that we are limited by the lack of a dataset with a ground truth other than the IRA dataset. 

For future work, we aim to extend our model to utilize derived attributes such as: sentiments, emotions, and topics using the content of the posts. Investigating methods to automatically determine time intervals will also be important for real-time data. Further, we intend to work on a case study to use this approach on other social media datasets.

\bibliographystyle{ieeetr}
\bibliography{reference}

\end{document}